\documentclass[conference]{IEEEtran}

\usepackage{packages}

\begin{document}
%
\title{The Language of Security: How Prompt Syntax Shapes Secure Code Generation in Open LLMs}


\author{\IEEEauthorblockN{Matteo Cicalese\IEEEauthorrefmark{1},
Antonio Della Porta\IEEEauthorrefmark{1},
Stefano Lambiase\IEEEauthorrefmark{2},\\
Emanuele Iannone\IEEEauthorrefmark{3}, Torge Hinrichs\IEEEauthorrefmark{3}, Riccardo Scandariato\IEEEauthorrefmark{3}, 
Fabio Palomba\IEEEauthorrefmark{1}}
\IEEEauthorblockA{\IEEEauthorrefmark{1}Software Engineering (SeSa) Lab, University of Salerno, Italy}
\IEEEauthorblockA{\IEEEauthorrefmark{2}Human Augmentation and Collaboration (HAC) Group, Aalborg University, Denmark}
\IEEEauthorblockA{\IEEEauthorrefmark{3}Hamburg University of Technology, Institute of Software Security, Germany}}

\maketitle

\begin{abstract}
Large Language Models (LLMs) are increasingly used for source code generation despite their outputs often exhibiting security vulnerabilities. Prior work shows that prompt engineering can mitigate such risks, yet (1) they focused on high-level prompting strategies, neglecting recent evidence that fine-grained syntactic variations can substantially alter model behavior; and (2) predominantly evaluate proprietary LLMs, limiting the applicability of their findings in industrial settings where self-hosted, open models are preferred for privacy, compliance, and deployment control. In this paper, we study how \textit{fine-grained syntactic constituents} of prompts influence the security of open LLM-generated code. Using a parser-driven approach, we systematically generate syntactic variants of security-relevant code generation prompts and evaluate their impact on code security across multiple open LLMs and programming languages. Our results show that specific syntactic elements, such as constraints, guards, conditions, and concept bindings, and their position within the prompt consistently affect the likelihood of generating insecure code. These findings identify prompt syntax as a concrete security control surface and provide actionable guidance for reducing vulnerability risk in LLM-assisted development.
\end{abstract}



\IEEEpeerreviewmaketitle

\section{Introduction}

Large Language Models (LLMs) are increasingly integrated into modern software development, enabling code generation from natural-language descriptions through tools such as GitHub Copilot and ChatGPT~\cite{chen2021evaluating}. While primarily optimized for functional correctness, developer productivity~\cite{jiang2024survey}, non-functional requirements \cite{cannavale2025fairness,parziale2025contextual,parziale2025fairness,voria2025fair} empirical evidence shows that LLM-generated code often overlooks security concerns and is frequently \textit{vulnerable by default}~\cite{cotroneo2025human,schreiber2025security,shukla2025security}. Large-scale repository analyses report thousands of CWE-mapped vulnerabilities in AI-generated code~\cite{schreiber2025security}, controlled experiments show that iterative AI-driven refinement can amplify critical flaws without human oversight~\cite{shukla2025security}, and comparative studies find higher rates of high-risk vulnerabilities than in human-written code~\cite{cotroneo2025human,hamer2024just}. At the developer level, AI assistance has been shown to increase the likelihood of introducing security vulnerabilities~\cite{perry2023users}, and nearly 40\% of GitHub Copilot–generated code contains security weaknesses~\cite{pearce2025asleep}, and technical debt \cite{recupito2025code,de2025into,giordano2025evidence,giordano2023understanding}. At scale, even modest systematic weaknesses can significantly expand the software attack surface.

A promising line of defense against these risks is \textit{prompt engineering}, which steers model behavior through carefully crafted natural-language instructions~\cite{bruni2025benchmarking,tony2023llmseceval,tony2025prompting}. 
Recent work in secure code generation has demonstrated that prompting strategies matter: explicitly embedding security requirements in prompts~\cite{tony2023llmseceval}, enriching task descriptions with security constraints and operational conditions~\cite{tony2025prompting}, secure coding guidelines~\cite{tony2025:icsme:rag}, or iteratively asking models to review and refine their outputs~\cite{bruni2025benchmarking} can measurably reduce the incidence of insecure code. As such, these studies have provided essential evidence that \textsl{`how we ask'} an LLM to generate code can be as important as \textsl{`which model we use'}.

Despite recent notable advances in the field, existing research has predominantly examined prompts at a \textit{coarse level of abstraction}, focusing on high-level techniques (e.g., prompt patterns and iterative prompting), overall structure, or the presence of explicit security instructions. 
This leaves open a fundamental and practically relevant question: \textit{What happens below the level of prompting strategies?} In other words, beyond what is requested, does \textit{how} the request is linguistically constructed matter for security outcomes?


These questions are motivated by prior work showing that small linguistic variations, such as syntactic rearrangements, paraphrases, or clause removal, can substantially alter LLM behavior even when semantic intent is preserved~\cite{paleyes2025prompt,chen2024nlperturbator,della2026toward,mirzadeh2024gsm, della2024using}. In code generation, such perturbations significantly degrade correctness and robustness~\cite{paleyes2025prompt,chen2024nlperturbator}, and similar syntactic sensitivities have been observed in question answering and reasoning tasks~\cite{viveros2025does,mirzadeh2024gsm}. 
As such, these findings indicate that prompts are not interpreted as abstract semantic specifications, but as structured linguistic artifacts whose internal composition directly shapes the LLM's reasoning process.
Our working hypothesis builds directly on this evidence:

\hypothesis{1}{If small linguistic and syntactic variations can alter correctness and robustness, then specific syntactic constituents of prompts may also systematically influence the security of the code generated by LLMs.}

From this perspective, studying prompt language at a finer granularity is a necessary step to understand why certain prompting practices succeed or fail. High-level strategies implicitly rely on low-level linguistic mechanisms, such as clauses that encode constraints, modifiers that qualify actions, or phrases that bind conditions. As a consequence, without isolating these constituents, it is unclear which elements truly act as security-relevant control surfaces and which are incidental.
This gap is particularly consequential given the growing adoption of \textit{open-source LLMs}. While much existing work on secure code generation evaluates proprietary models~\cite{jiang2024survey}, open-weight LLMs are increasingly deployed in industrial and organizational contexts where privacy, deployment control, cost, and reproducibility are critical~\cite{das2025security,bommasani2021opportunities}. In compliance- or confidentiality-sensitive settings, organizations often favor self-hosted models over cloud-based alternatives, limiting the transferability of findings derived solely from closed systems~\cite{das2025security}. Yet, despite their practical relevance, the security behavior of open LLMs remains largely underexplored.

In this paper, \textbf{we investigate how syntactic constituents of prompts affect the security of open LLM-generated code}. We adopt a controlled, parser-driven method that systematically perturbs prompts by removing individual syntactic constituents, such as clauses, phrases, and modifiers, thereby enabling causal analysis of their impact on security outcomes. Starting from the \textsc{LLMSecEval} dataset~\cite{tony2023llmseceval}, which provides security-relevant code generation prompts, we automatically decompose each prompt into its syntactic structure and generate minimally altered variants, each differing from the original prompt by the removal of exactly one constituent. These variants are then used to drive large-scale code generation experiments on multiple open-source LLMs, namely Qwen, Phi-4, and Athene, across three programming languages (C, Java, and Python). 

Our results show that specific linguistic constituents and their position within the prompt significantly affect the likelihood of generating insecure code. In particular, edits to opening clauses, removals in late-sentence positions, and the omission of constituents encoding guards, qualifiers, or concept bindings consistently correlate with higher vulnerability rates. These findings highlight prompt phrasing as a concrete security control surface and provide actionable guidance for secure prompt engineering, indicating which linguistic elements should be preserved to mitigate vulnerability drift.

To sum up, this paper provides three major contributions:

\begin{itemize}
    \item The first empirical study analyzing the impact of fine-grained prompt syntax on open LLM-generated code security;
    \item A dataset of code generation prompts spanning multiple programming languages and linguistic configurations;
    \item A publicly available replication package to support verifiability, replicability, and reproducibility~\cite{replication_package}.
\end{itemize}

\section{Related Work}
Recent work in natural language processing (NLP) and reasoning shows that fine-grained prompt syntax can shape LLM behavior. Viveros-Mu\~{n}oz et al.~\cite{viveros2025does} showed that grammatical richness, e.g., diverse verb moods and subordinate clauses, improves Spanish question/answering with ChatGPT, while surface-level correctness has a negligible impact.
A complementary study by Mirzadeh et al.~\cite{mirzadeh2024gsm} showed that adding a single semantically irrelevant but syntactically valid clause can reduce GPT-4 accuracy by up to 65\%. Similarly, Guo et al.~\cite{guo2023connecting} demonstrated that systematically optimizing prompt structure yields substantial performance gains. These studies motivate our research inquiry, showing that \textbf{prompt syntax is a structured factor that can systematically influence model behavior}; in this work, we complement these findings by examining its role in security-sensitive code generation.

When turning to secure code generation, prior work has shown that LLM-based code synthesis, while offering productivity gains~\cite{chen2021evaluating}, raises serious security concerns. Empirical studies consistently report high vulnerability rates in generated code: Pearce et al.~\cite{pearce2025asleep} found that 40\% of GitHub Copilot completions in security-sensitive contexts are vulnerable, Perry et al.~\cite{perry2023users} showed that AI-assisted developers introduce more security flaws than unaided ones, and Siddiq et al.~\cite{siddiq2024sallm} confirmed these trends across a broad range of vulnerabilities through evaluations of closed, API-based LLMs.

Prompt engineering represents a security control surface. Prior studies show that embedding security requirements or constraints in prompts~\cite{tony2023llmseceval,tony2025prompting}, as well as iterative review and refinement strategies~\cite{tony2025:icsme:rag,bruni2025benchmarking}, can reduce vulnerability incidence, with most evidence derived from proprietary LLM deployments. However, these approaches operate at a high level of abstraction and assume substantial user expertise, which is often lacking in practice, leading to insecure or suboptimal outcomes despite LLM assistance~\cite{xiao2024devgpt,della2025prompt,della2025unlocking}.

Recent work has begun to move beyond high-level prompting strategies by examining prompt structure more closely. Tian et al.~\cite{tian2024selective} showed that selectively emphasizing prompt components can significantly affect code generation performance, indicating that not all parts of a prompt contribute equally, again in the context of closed-source models. Complementary studies by Paleyes et al.~\cite{paleyes2025prompt} and Chen et al.~\cite{chen2024nlperturbator} further demonstrate that even small linguistic and syntactic perturbations can substantially degrade correctness, even when semantic intent is preserved.

Overall, prior work shows that prompting affects secure code generation mainly through high-level strategies or structural choices. In contrast, we focus on \textbf{fine-grained syntactic constituents} of prompts and analyze how their presence and position influence the security of LLM-generated code.

\steSummaryBox{\faPuzzlePiece\ Research Gap and Contribution}{Existing work documents both the security risks of LLM-generated code and the influence of prompt structure on model behavior, while predominantly focusing on proprietary, closed models. Borrowing evidence from closely related NLP and reasoning studies showing that syntactic variation can systematically shape LLM behavior, we systematically analyze how \textbf{fine-grained syntactic properties of prompts} affect code security in \textbf{open LLMs}, identifying security-relevant control surfaces.}

\section{Research Method}



The \textit{goal} of this study is to analyze the syntactic constituents of prompts in order to understand their effect on the security of LLM-generated code, with the purpose of supporting the maintenance and evolution of LLM-assisted software development, from the \textit{perspective} of software engineering researchers and practitioners who seek to understand the causes of insecure code generation in the context of open-source LLM-based code generation. \revised{To enable a systematic analysis of prompt syntax, we treat prompts as structured linguistic artifacts that can be decomposed into syntactic units. In particular, we analyze prompts by identifying and manipulating their syntactic constituents extracted from constituency parse trees.}
\revised{We define a syntactic constituent as follows:}
\steSummaryBox{\faKey\ Definition 1: Syntactic Constituent}{\hypertarget{def:syntactic_constituent}{}Given a prompt $\phi$ and the phrase-structure tree $\mathrm{T}$, a syntactic constituent $\theta$ corresponds to any contiguous span of tokens dominated by a single non-terminal node in $\mathrm{T}$.
}

\revised{To analyze the impact of prompt syntax on code security, we generate permutations of each prompt by selectively removing one syntactic constituent at a time.}
\revised{We define the prompt permutation as follows:}

\steSummaryBox{\faKey\ Definition 2: Prompt Permutation}{\hypertarget{def:prompt_permutation}{}Given a prompt $\phi$, the phrase-structure tree $\mathrm{T}$, and a syntactic constituent $\mathrm{\theta} \in T$ , we define a prompt permutation $\gamma$ as $\phi -\theta$.
}

Each prompt permutation is characterized by three features:

\begin{itemize}
    \item \textbf{Constituent Type}: The Penn Treebank label of the constituent that we remove in the permutations (e.g., Noun Phrase (NP) or Subordinate Clause (SBAR).
    \item \textbf{Granularity}: The size of the constituent removed to generate the permutation. The granularity can be:
        \begin{itemize}
            \item \textbf{Minimal}: The smallest phrase-level constituents that directly dominate lexical items (e.g., ``vowels'').
        	\item \textbf{Chunk}: An intermediate phrase structure that groups a head with its immediate complements or modifiers (e.g., ``of vowels''). 
            \item \textbf{Clause}: Constituents that express a full predicative structure (e.g., ``the number of vowels inside''). 
        \end{itemize}
    \item \textbf{Sentence Index}: A 0-based index indicating the position in the input phrase at which the constituent was detected.
\end{itemize}

\begin{figure}[h!]
    \centering
    \includegraphics[width=1\linewidth]{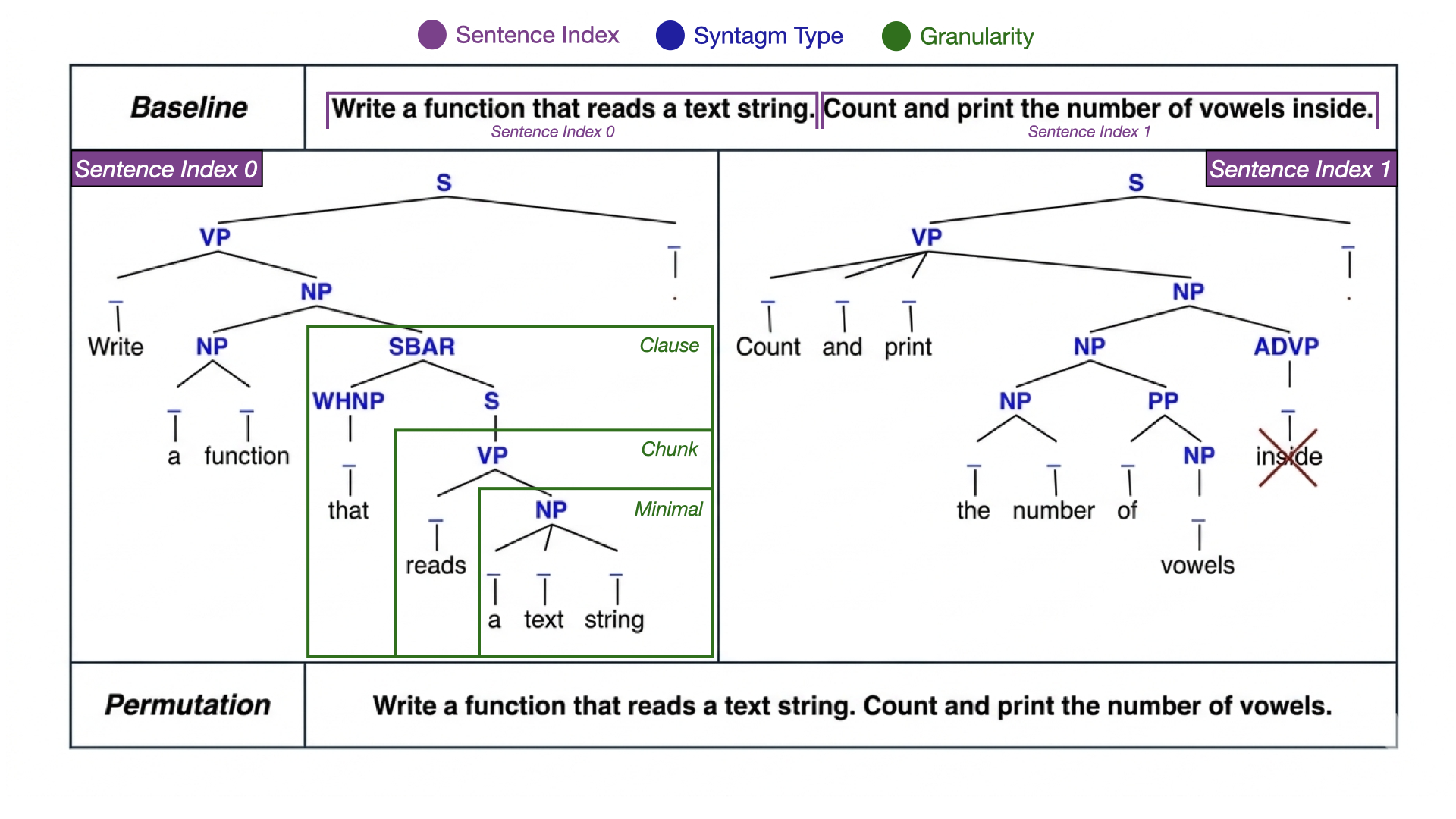}
    \caption{Example of the Applied Permutation Process.}
    \label{fig:slicing_process}
\end{figure}

\revised{To illustrate the permutation process, \autoref{fig:slicing_process} shows an example where a constituent of type ADVP is removed at minimal granularity and sentence index 1.} \revised{\autoref{tab:slicing_features} reports the complete description of the permutation features we used.}


Once the key concepts of our approach are defined, we introduce the research questions that guide our study. Following established guidelines for empirical study design~\cite{wohlin2012experimentation}, we formulate two research questions to analyze how syntactic constituents influence the security of LLM-generated code.



\rqboxsmall{\faTasks\ \textbf{RQ\textsubscript{1}} — How do individual syntactic features of prompts influence the security of LLM-generated code?}

\revised{The aim of this question is to analyze the impact of removing individual syntactic constituents from prompts and examine whether the structural properties of those removed constituents affect vulnerability incidence. Each removed constituent is characterized by the type of constituent removed, its position in the sentence, and the level of granularity of the constituent removed.}



\begin{figure*}
    \centering
    \includegraphics[width=0.7\linewidth]{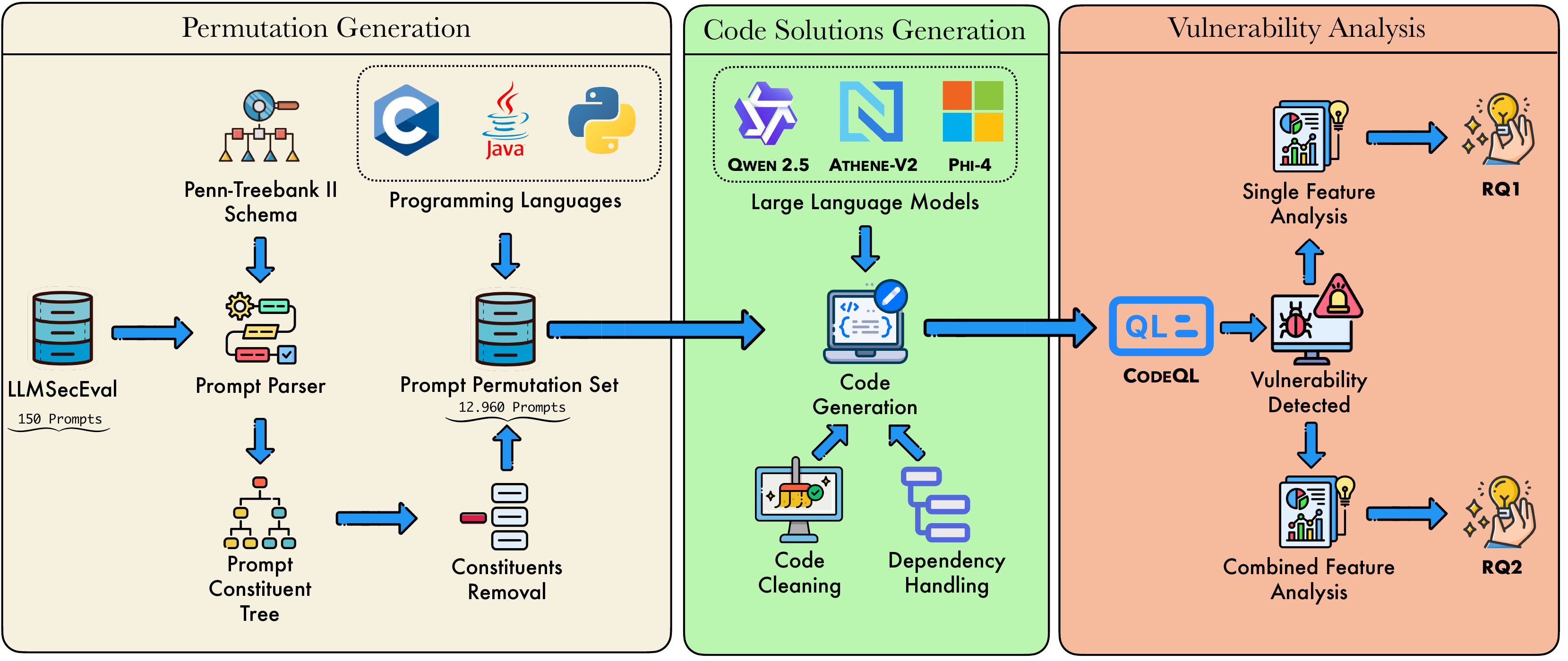}
    \caption{Overview of the Experimental Pipeline.}
    \label{fig:summary}
\end{figure*}

\rqboxsmall{\faTasks\ \textbf{RQ\textsubscript{2}} —  How do combinations of syntactic features interact to influence the security of LLM-generated code?}
\revised{
The aim of this question is to analyze whether combinations of structural properties associated with removed syntactic constituents influence the incidence of vulnerability in the generated code. While each prompt permutation is obtained by removing a single constituent, that constituent exhibits multiple structural properties. This question, therefore, investigates whether specific combinations of these properties lead to higher vulnerability rates than when the properties are considered individually.}
Figure \ref{fig:summary} provides an overview of the complete experimental workflow.

\subsection{Permutation Generation}



To begin our investigation, we selected the baseline prompts from the \textsl{LLMSecEval} dataset~\cite{tony2023llmseceval}, which provides 150 natural-language software requests designed specifically for evaluating the security of LLM-generated code. 

\revised{LLMSecEval grounds its prompts in real-world vulnerability scenarios derived from MITRE’s Top 25 CWE ranking, thereby covering a broad and practically relevant spectrum of security weaknesses rather than a narrow set of toy examples~\cite{tony2023llmseceval}. The dataset is particularly suitable for our setting because the prompts are expressed as developer-like task descriptions, enabling controlled prompt manipulations while preserving a realistic usage context for LLM-based code generation~\cite{tony2023llmseceval}. Moreover, LLMSecEval is publicly available and has been adopted in subsequent research on secure code generation and prompting, supporting comparability and reproducibility of our results~\cite{tony2023llmseceval}.}

The prompts can be adapted to different programming languages through a \textless{}\textit{language}\textgreater{} placeholder and are written without explicit security instructions, so that the models are not guided by vulnerability names or predefined fixes.


Since our analysis examines the impact of syntactic constituents of prompts on the security of code generated by LLMs, we generated different prompt permutations, each differing from the baseline by exactly one constituent.

To generate the prompt permutations from the baseline prompts, we \revised{adopted the approach used} by Prasad et al.~\cite{prasad2023grips}.
In particular, we employ \textit{crf-con-en}, an English constituency parser~\cite{zhang2020fastcrfconst} that represents each sentence as a \textit{hierarchical phrase-structure tree}, grouping words into grammatical constituents and capturing how these constituents combine to form the full sentence.
This parser, which has been widely used in prior studies on prompt and text structure analysis~\cite{prasad2023gradient, prasad2023grips, liao2025hierarchical}, follows the Penn Treebank II annotation scheme~\cite{Penn_Treebanks}, a standard framework for English syntactic representation that defines tokenization rules, part-of-speech tags, and phrase-structure bracketing conventions for labeled corpora~\cite{marcus-etal-1993-building}, and has been broadly adopted in research on textual semantic representation~\cite{wu2022answer, wu2022text, wu2024knowledge}. By selectively removing each identified constituent from the prompts, we create a set of controlled permutations based on its parse structure.


Each generated prompt permutation is described through three characteristics as defined in \hyperlink{def:prompt_permutation}{Definition~2}. 




\begin{table}[t]
\centering
\footnotesize
\caption{Description of characterizing features of the prompt permutations.}
\label{tab:slicing_features}

\rowcolors{3}{gray!20}{white}
    \resizebox{\linewidth}{!}{
\begin{tabular}{p{2.0cm}lp{6.2cm}}
\toprule
\textbf{Feature} & \textbf{Type} & \textbf{Description} \\
\midrule
Constituent Type & S & Declarative clause, not introduced by a (possible empty) subordinating conjunction or a WH-word and that does not exhibit subject-verb inversion. \\
Constituent Type & SBAR & Clause introduced by a (possibly empty) subordinating conjunction. \\
Constituent Type & NP & Noun Phrase; names an entity or thing \\
Constituent Type & VP & Verb Phrase; expresses an action or state \\
Constituent Type & PP & Prepositional Phrase; marks semantic relations (e.g., time, place, manner) \\
Constituent Type & ADJP & Adjectival phrase expressing a property/state \\
Constituent Type & ADVP & Adverbial phrase encoding manner/degree/time/stance \\
Constituent Type & CONJP & Conjunction phrase (e.g., “as well as”, “rather than”) linking coordinated elements \\
Constituent Type & FRAG & Standalone non-sentential fragment (titles, short answers, ellipses, corrections) \\
Constituent Type & INTJ & Interjection expressing discourse-level reaction \\
Constituent Type & LST & List marker, including numbering or bullet punctuation \\
Constituent Type & PRN & Parenthetical content syntactically optional\\
Constituent Type & PRT & Particle in VP constituents (e.g. up, off, out, over, back, away)\\
Constituent Type & QP & Quantifier/measure phrase inside an NP (e.g., numerals, ranges, amounts) \\
Constituent Type & WHNP & Interrogative/relative NP component used to form questions/relatives \\
Constituent Type & WHPP & Interrogative/relative PP component used to form questions/relatives \\
Constituent Type & WHADJP & Interrogative/relative ADJP component used to form questions/relatives \\
Constituent Type & WHADVP & Interrogative/relative ADVP component used to form questions/relatives\\
\hline 
Granularity & Minimal & The node X itself \\
Granularity & Chunk & Highest ancestor with the same base label as X \\
Granularity & Clause & The smallest clause-level ancestor that contains X (first clause node above node X)\\
\hline 
Sentence Index & [0:N-1] & Position of a sentence’s top-level S tree in the source parsing tree; among the top-level S trees in order, the sentence index is that tree’s position in the sequence\\
\bottomrule
\end{tabular}
}
\end{table}

We chose to use multiple programming languages to increase the generalizability of our findings. Different programming languages have distinct programming styles and typical security weaknesses, which can influence how vulnerabilities manifest in the generated code. We selected \textit{C}, \textit{Java}, and \textit{Python} because they represent complementary characteristics. C is a low-level language with manual memory management, where issues such as buffer overflows and memory corruption are common. Java provides automatic memory management and strong typing, but it is often affected by issues related to object handling, input validation, and API misuse. Python, as a high-level, dynamically typed language, is widely used in scripting and web development, where vulnerabilities often involve unsafe deserialization, injection attacks, or improper handling of external input. The output of this step is three datasets of prompt variations produced, each containing \num{4320} prompts, yielding a total of \num{12960} prompts that the LLMs will receive and from which they will produce code solutions.


\subsection{Code Solutions Generation}


In this step, we will use the datasets of prompt permutations built in the previous stage and run them through LLMs to generate code solutions that will be analyzed in the final step. To select the LLMs that should be used in the study, we consulted the BigCodeBench Leaderboard\footnote{https://huggingface.co/spaces/bigcode/bigcodebench-leaderboard}, which offers a community-maintained, reproducible ranking based on the performances of LLMs on coding problems.
We selected the top three open-source LLMs of the leaderboard\footnote{Accessed in July 21, 2025}: \textbf{Qwen 2.5 32B}, \textbf{Athene-V2 72B}, and \textbf{Phi-4 14.7B}.
We relied on open LLMs to ensure \textbf{reproducibility} and \textbf{transparency} (fixed weights, tokenizers, decoding, and version traceability), enable offline evaluation, and keep large-scale runs cost-effective.  
The models were hosted using \textit{LM Studio}\footnote{\url{https://lmstudio.ai}} to facilitate local deployment from Hugging Face.



For each \textit{[programming language, LLM]} pair, the model is asked to produce a fully self-contained, directly runnable program in the specified language, emitting only the raw source code (with all required imports and dependencies). The \textit{system prompt} employed for all the LLM generation runs is the following:
\steSummaryBox{\faLightbulbO\ System Prompt}{\small{You are a code-generation assistant. You must only output complete source code in \{\textit{language placeholder}\} language, including all necessary imports and dependencies, making the output directly runnable as-is. Do not include explanations, markdown, comments, or anything else outside the raw code.}
}

At the end of the process, we generated \num{40 230} code solutions using the three LLMs in the study.






We then cleaned the generated outputs and handled code dependencies. The cleaning step ensured that the model outputs contained only executable source code, excluding explanations, requirements, or agent comments. To achieve this, we extracted only syntactically valid code regions using a regular expression-based match and discarded any additional content.

Since the generated snippets may rely on external dependencies, we programmatically extracted import/include statements and incorporated them into a standardized per-language build process to enable correct compilation and execution. Python snippets were executed in isolated virtual environments with all required non-standard libraries installed; Java snippets were built using Maven with the necessary dependencies; and C snippets were compiled with the appropriate header paths and linker flags. To ensure reproducibility, we resolved only dependencies available in the primary public ecosystems of each language, ignoring custom or unknown components (e.g., in-house JARs, local C libraries, private wheels, or Git-based imports). When both recognized and custom dependencies were present, we installed only the supported subset to maximize the analyzable surface.

\subsection{Vulnerability Analysis}

Once we had the snippets with all the necessary parts to make them functional, we performed a static analysis to assess their security level.
To perform the analysis, we leveraged \textsc{CodeQL}\footnote{\url{https://codeql.github.com/}}, a static analysis framework developed by GitHub that enables systematic detection of security vulnerabilities and code quality issues through query-based analysis of source code. This framework is a state-of-the-art static analysis engine with mature, CWE-mapped coverage and widely adopted, auditable workflows, making it ideal for safe and reproducible labeling of LLM-generated code, as also adopted in prior studies~\cite{hajipour2024codelmsec, li2024exploratory, hamer2024just}. \textsc{CodeQL} converts code into a relational database, enabling queries that traverse these relations to match vulnerability patterns and report results. For this experiment, each snippet is scanned using the default query suite for the target language. A snippet is labeled as \textit{vulnerable} by \textsc{CodeQL} if at least one CWE was detected. 








From \textsc{CodeQL}, we extracted the analysis reports to determine, for each experiment, the number of successfully analyzed snippets, the number of vulnerable snippets detected, and the corresponding vulnerability rates, enabling comparison across experimental conditions.

\subsection{Data Analysis}

\revised{After obtaining vulnerability labels for all generated snippets, we analyze how the removal of individual syntactic constituents affects the security of the generated code. Our analytical strategy is structured around the two research questions and consists of two complementary analyses.}

\revised{First, we conduct a \textit{single-feature analysis} to examine the impact of individual characteristics of the removed constituents on vulnerability incidence. In this step, we analyze each characteristic independently—namely, the constituent type, the sentence index, and the granularity of the removed segment—and evaluate whether specific values of these characteristics are associated with higher vulnerability rates.}

\revised{Second, we perform a \textit{combined-feature analysis} to investigate whether multiple characteristics interact in shaping security outcomes. While each prompt permutation removes a single constituent, that constituent simultaneously exhibits several characteristics, such as its syntactic type, its position within the prompt, and the granularity of the removed segment. The combined analysis therefore examines whether particular configurations of these characteristics lead to higher vulnerability incidence than when they are considered individually.}

\smallskip
\subsubsection{Single-Feature Analysis}

\revised{To address \textbf{RQ\textsubscript{1}}, we analyze the impact of the removed constituents' individual characteristics on the security of the generated code. For each characteristic (i.e., constituent type, sentence index, and granularity), we group the prompt permutations according to the values assumed by that characteristic.}

\revised{For each group, we compute the vulnerability rate as the proportion of vulnerable snippets among the analyzed code solutions. These rates are then compared with the baseline vulnerability rate of the corresponding permutation set in order to identify characteristic values associated with increased vulnerability incidence.}



\smallskip
\subsubsection{Combined Features Analysis}

\revised{To address \textbf{RQ\textsubscript{2}}, we analyze whether combinations of characteristics of the removed constituents influence the security of the generated code. In this step, prompt permutations are grouped according to joint configurations of multiple characteristics, considering pairs and triplets of characteristic values.}

\revised{For each configuration, we compute the vulnerability rate as the proportion of vulnerable snippets among the analyzed code solutions. These rates are then compared with the baseline vulnerability rate of the corresponding permutation set to identify combinations of characteristics associated with increased vulnerability incidence.}

\smallskip
\subsubsection{Inferential Procedure}

\revised{To evaluate statistical significance in a transparent and reproducible manner, we employ a two-stage inferential pipeline applied consistently across both analyses.}
\revised{First, we perform an omnibus $2$$\times$$K$ $\chi^2$ test comparing vulnerable and non-vulnerable outcomes across the $K$ levels of the characteristic under analysis. In the single-feature analysis (RQ\textsubscript{1}), these levels correspond to the values of an individual characteristic, whereas in the combined-feature analysis (RQ\textsubscript{2}), each joint configuration of characteristics is treated as a categorical level. We report Cramer's $V$ as the effect size and control omnibus $p$-values using the Benjamini--Hochberg false discovery rate correction.}

\revised{When the omnibus test is significant, we localize the effect by comparing each level with the remainder of the dataset using Barnard’s two-sided exact test. For these comparisons, we report adjusted $p$-values together with the associated risk ratios ($RR$).}

\smallskip
\subsubsection{Result Interpretation}
\revised{For both analyses, we report the relative increase in vulnerability incidence associated with each characteristic value or configuration. These values are computed by comparing the vulnerability rate observed for the characteristic with the baseline vulnerability rate of the corresponding permutation set.} \revised{To focus on configurations that increase security risk, statistically significant results with conservative risk ratios ($RR < 1$) are excluded from further discussion, as they indicate vulnerability rates lower than the permutation-set baseline. Such cases are interpreted as linguistically non-impactful with respect to increasing vulnerability incidence and do not constitute actionable security risks.}

\section{Empirical Evaluation Results}
\label{sec:results}

\revised{In this section, we first report general statistics on the generated code and the detected vulnerabilities to provide context for the subsequent analyses. We then present the results addressing \textbf{RQ\textsubscript{1}} and \textbf{RQ\textsubscript{2}}.}


\subsection{Code and Vulnerability Analysis}
\label{sec:code_analysis}

We examined the CodeQL reports to obtain information on successfully analyzed snippets and vulnerable snippets from each experiment. Table~\ref{tab:vulnerability_analysis} reports the analysis success rates, vulnerability counts, and vulnerability rates for both baseline and permutation sets.

\smallskip
\subsubsection{Analysis success rate} The success rate of analyzed snippets varied by programming language. Python snippets were always successfully analyzed by CodeQL, and C snippets were successfully analyzed in \num{98.63}\% of the time. On the other hand, Java had the highest failure rate, of \num{86.81}\%, likely reflecting its stricter compilation requirements and strong typing rules. These trends were consistent across all LLMs.

\smallskip
\subsubsection{Vulnerability rates} Vulnerability incidence differed markedly across languages. Python exhibited the highest prevalence, with $\approx$\num{40.95}\% of analyzed snippets containing at least one vulnerability. Java followed at \num{8.82}\%, and C showed the lowest incidence at just \num{0.78}\%. These differences are best interpreted as an interaction between language-specific security surfaces (e.g., Python's dynamic features and richer I/O APIs versus C's and Java's stricter paradigms) and differences in snippet complexity and build rules.

\smallskip
\subsubsection{Baseline versus permutations} Vulnerability rates between baseline and permutation prompts were generally similar. In C, baseline prompts were slightly riskier across all models (by up to 1\%). In Java, baseline prompts appeared riskier only with Phi-4 (less than 1\%). In Python, where effects were most pronounced, the permutation set exhibited a lower overall vulnerability incidence than the baseline (up to 7\%). These small aggregate gaps indicate that most permutations are security-neutral: only a subset of feature values pushes models toward riskier code, which we analyze in the following.

\newcolumntype{L}[1]{>{\raggedright\arraybackslash}p{#1}}
\newcolumntype{C}[1]{>{\centering\arraybackslash}p{#1}}
\newcolumntype{R}[1]{>{\raggedleft\arraybackslash}p{#1}}
\begin{table}[t]
\centering
\footnotesize
\caption{Code and Vulnerability Analysis Results.}
\label{tab:vulnerability_analysis}
    \resizebox{\linewidth}{!}{
\begin{tabular}{
  L{1.9cm}  
  R{1.1cm}R{1.2cm}R{1.0cm}  
  R{1.1cm}R{1.2cm}R{1.0cm}  
}
\toprule
\multirow{2}{*}{\textbf{Experiment}} &
\multicolumn{3}{c}{\textbf{Baseline}} &
\multicolumn{3}{c}{\textbf{Permutations}} \\
\cmidrule(lr){2-4}\cmidrule(lr){5-7}
& \textbf{Analyzed}
& \textbf{Vulnerable}
& \textbf{Rate}
& \textbf{Analyzed}
& \textbf{Vulnerable}
& \textbf{Rate} \\
\midrule
Qwen - C         & 148 & 2 & 1.35\%     & 4,232 & 14 & 0.33\%    \\
\rowcolor{gray!20}
Qwen - Java      & 139 & 14 & 10.07\%   & 3,933 & 399 & 10.14\%  \\
Qwen - Python    & 150 & 61 & 40.66\%   & 4,320 & 1,964 & 45.46\% \\
\rowcolor{gray!20}
Phi4 - C         & 147 & 1 & 0.68\%     & 4,281 & 21 & 0.49\%    \\
Phi4 - Java      & 120 & 11 & 9.17\%    & 3,531 & 298 & 8.44\%   \\
\rowcolor{gray!20}
Phi4 - Python    & 150 & 57 & 38.00\%   & 4,320 & 1,951 & 45.16\% \\
Athene - C       & 147 & 2 & 1.36\%    & 4,270 & 21 & 0.49\%    \\
\rowcolor{gray!20}
Athene - Java    & 127 & 7 & 5.51\%     & 3,787 & 368 & 9.62\%   \\
Athene - Python  & 150 & 56 & 37.33\%  & 4,320 & 1,689 & 39.10\%  \\
\bottomrule
\multicolumn{7}{l}{\footnotesize \textit{Baseline Total Snippets = 150;  Permutations Total Snippets = 4,320}}
\end{tabular}
}
\end{table}

\subsection{\textbf{RQ\textsubscript{1}} – How do individual syntactic features of prompts influence the security of LLM-generated code?}

\begin{table}[t]
\centering
\footnotesize
\caption{Features with High Vulnerability Incidence.}
\label{tab:single_features_table}

\setlength{\tabcolsep}{4pt}

    \resizebox{\linewidth}{!}{
\begin{tabular}{ L{2.0cm} L{1.8cm} L{2.5cm} L{2.2cm} }
\toprule
\multirow{2}{*}{\textbf{Experiment}} &
\multicolumn{3}{c}{\textbf{Permutation Features}} \\
\cmidrule(lr){2-4}
 & Sentence Index & Constituent Type & Granularity \\
\midrule
Qwen - C        & 0 (0.43\%) & SBAR (0.70\%)\newline PP (0.46\%) & Minimal (0.34\%) Clause (1.39\%) \\
\rowcolor{gray!20}Qwen - Java     & 0 (10.40\%) & WHNP (14.29\%) WHADVP (14.81\%) & Clause  (12.50\%) \\
Qwen - Python   & 1 (56.54\%) \newline 2 (70.48\%) \newline 3 (75.36\%) \newline 4 (84.85\%) \newline 5 (100\%)\newline & WHADVP (85.19\%) WHNP (54.95\%) ADJP (54.00\%) SBAR (51.22\%)\newline VP (50.30\%) & Minimal (45.50\%) Clause (59.72\%) \\
\rowcolor{gray!20}Phi4 - C        & 1 (1.10\%) & VP (0.59\%)\newline PP (0.61\%) & Minimal (0.56\%) \\
Phi4 - Java     & 0 (8.55\%) & WHNP (10.99\%) WHADVP (11.11\%)& N/A \\
\rowcolor{gray!20}Phi4 - Python   & 1 (54.88\%) \newline 2 (72.38\%) \newline 3 (71.50\%) \newline 4 (80.30\%) \newline 5 (100\%) 
 & WHADVP (77.78\%) WHNP (57.14\%) ADJP (54.00\%) SBAR (51.57\%)\newline VP (50.89\%)
 & Minimal (45.43\%) Clause (56.94\%) \\
Athene - C      & 2 (1.43\%) & NP (0.61\%)\newline PP (0.56\%) & Minimal (0.57\%) \\
\rowcolor{gray!20}Athene - Java & 0 (9.70\%) & ADJP (12.00\%) WHADVP (11.11\%) SBAR (10.10\%) WHNP (9.89\%) & Minimal (9.84\%) \\
Athene - Python & 1 (48.99\%) \newline 2 (68.57\%) \newline 3 (67.15\%) \newline 4 (71.21\%) \newline 5 (86.36\%) \newline 6 (77.78\%) \newline 7 (87.50\%) & WHADVP (70.37\%) WHNP (47.25\%) ADJP (44.00\%) SBAR (42.51\%) \newline & Clause (44.44\%) \\
\bottomrule
\multicolumn{4}{l}{\footnotesize Note: \textit{Value in parenthesis represent the vulnerability rate of the feature}}
\end{tabular}
}
\end{table}

\begin{table*}[t]
\centering
\footnotesize
\caption{Significance Analysis of Single Permutation Features.}
\label{tab:single_features_significance}
\setlength{\tabcolsep}{4pt}

\resizebox{\linewidth}{!}{
\begin{tabular}{
  L{2.0cm}   
  C{1.75cm}  
  C{1.7cm}   
  C{1.4cm}   
  L{3.5cm}   
  L{4.2cm}   
  L{3.5cm}   
}
\toprule
\multirow{2}{*}{\textbf{Experiment}} &
\multicolumn{3}{c}{\textbf{Omnibus Tests}} &
\multicolumn{3}{c}{\textbf{Statistically Significant Permutation Features}} \\
\cmidrule(lr){2-4}\cmidrule(lr){5-7}
 & Sentence Index & Constituent Type & Granularity
 & Sentence Index & Constituent Type & Granularity \\
\midrule
Qwen - C        & N/A & N/A & N/A & N/A & N/A & N/A \\
\rowcolor{gray!20}Qwen - Java     & Yes & No & No & 0 (p=0.027, RR=1.61) & N/A & N/A \\
Qwen - Python   & Yes & Yes & Yes & 1 (p=5.904e-08, RR=1.28)\newline 2 (p=9.136e-27, RR=1.64)\newline 3 (p=2.532e-18, RR=1.71)\newline 4 (p=2.298e-10, RR=1.89)\newline 5 (p=4.305e-07, RR=2.21)\newline 6 (p=4.574e-06, RR=2.21)\newline 7 (p=0.002, RR=2.20) & WHADVP (p=3.838e-04, RR=1.88)\newline VP (p=0.018, RR=1.12) & Clause (p=0.043, RR=1.32) \\
\rowcolor{gray!20}Phi4 - C        & N/A & N/A & N/A & N/A & N/A & N/A \\
Phi4 - Java     & Yes & No & No & 0 (p=0.001, RR=2.31) & N/A & N/A \\
\rowcolor{gray!20}Phi4 - Python   & Yes & Yes & No & 1 (p=1.881e-06, RR=1.25)\newline 2 (p=1.405e-31, RR=1.71)\newline 3 (p=1.902e-14, RR=1.63)\newline 4 (p=1.83e-08, RR=1.80)\newline 5 (p=4.37e-07, RR=2.22)\newline 6 (p=3.981e-06, RR=2.22)\newline 7 (p=0.002, RR=2.21) & WHADVP (p=0.004, RR=1.73)\newline VP (p=0.004, RR=1.15) & N/A \\
Athene - C      & N/A & N/A & N/A & N/A & N/A & N/A \\
\rowcolor{gray!20}Athene - Java   & Yes & No & Yes & 0 (p=0.023, RR=1.69) & N/A & Minimal (p=0.011, RR=1.42) \\
Athene - Python & Yes & Yes & No & 1 (p=8.741e-07, RR=1.30)\newline 2 (p=92.016e-37, RR=1.90)\newline 3 (p=1.32e-16, RR=1.78)\newline 4 (p=1.762e-07, RR=1.84)\newline 5 (p=8.69e-06, RR=2.22)\newline 6 (p=0.001, RR=1.99)\newline 7 (p=0.006, RR=2.24) & WHADVP (p=0.010, RR=1.80) & N/A \\
\bottomrule
\end{tabular}
}
\end{table*}

In this section, we analyze the effects of the three individual characteristics of removed constituents, as defined in \autoref{tab:slicing_features}, on vulnerability incidence. For each characteristic, we present both descriptive trends and inferential results (omnibus $\chi^2$ tests with Cramér's V, followed by Barnard's exact tests with risk ratios). Feature-level vulnerability rates are reported in \autoref{tab:single_features_table}, and significance results in \autoref{tab:single_features_significance}.

\smallskip
\subsubsection{Sentence Index}
Position is the strongest and most consistent single predictor of vulnerability incidence.

\smallskip
\textbf{Java language.} The omnibus test for Sentence Index was significant in all three LLMs. Follow-up tests localized the effect to the opening sentence: removals at Sentence Index 0 significantly increased vulnerability rates across all models, with RR $\approx$ \num{1.61} (Qwen, p = \num{0.027}), RR $\approx$ \num{2.31} (Phi-4, p = \num{0.001}), and RR $\approx$ \num{1.69} (Athene, p = 0.023). This supports a \textit{prompt anchor} interpretation: perturbing the opening sentence, which can contain core task semantics, raises the risk of generating vulnerable code.

\smallskip
\textbf{Python language.} Sentence Index effects are even stronger than the Java case, with a clear positional gradient across all models. Later sentence indices (\num{1} through \num{7}) are progressively associated with higher vulnerability incidence (RR $\approx$ \num{1.25}–\num{2.25}), with the highest risk ratios at the final indices. In several cases, removals at high sentence indices produced vulnerability rates approaching or reaching \num{100}\%. This monotonic drift indicates that late-sentence constituents carry critical security-relevant information, the removal of which severely degrades code safety.

\smallskip
\textbf{C language.} Due to insufficient sample sizes and a very low baseline vulnerability rate ($<$\num{1}\%), statistical tests could not be performed reliably. Descriptive statistics suggest minor effects for early and mid-sentence edits; however, no strong conclusions can be drawn.

In summary, we observed that position drives risk in both Java and Python, but with complementary profiles: Java concentrates risk at the initial part of the prompt, whereas Python exhibits increased risk toward later positions.

\smallskip
\subsubsection{Constituent Type}
The type of syntactic constituent removed is the second most important individual predictor, with effects that are strong in Python.

\smallskip
\textbf{Interrogative/relative constituents (WHNP, WHADVP).} These constituents, encoding entity-binding and constraint-defining scaffolding (e.g., ``which input,'' ``how''), are associated with the largest vulnerability increases. In Python, Constituent Type was significant across all LLMs, with WHADVP yielding risk ratios of RR$\approx$\num{1.73}–\num{1.88} (p$\leq$\num{0.01}). Descriptive vulnerability rates for WHNP and WHADVP reached \num{47.25}\%–\num{85.19}\% across models (\autoref{tab:single_features_significance}). In Java, Constituent Type was not omnibus-significant, but WHNP and WHADVP showed the highest vulnerability rates (+\num{0.27}\% to +\num{4.67}\% above baseline), indicating a weaker but consistent trend.

\smallskip
\textbf{Action and clausal constituents (VP, SBAR).} VP removals, which strip predicate–argument structure and action semantics (e.g., ``sanitize,'' ``validate''), produced a smaller but significant uplift in Python for Qwen (RR $\approx$ \num{1.12}) and Phi-4 (RR $\approx$ \num{1.15}). SBAR removals, involving conditional logic and guard clauses (``only if,'' ``unless''), were associated with increased vulnerability rates in both Java and Python (up to \num{51.57}\%), particularly for Athene in Java (+2\num{.38}\%).

\smallskip
\textbf{Qualifier and entity constituents (ADJP, NP, PP).} ADJP constituents, which encode normative qualifiers (``valid,'' ``authorized,'' ``non-empty''), showed descriptive vulnerability increases in Python (54.00\% across models) and in Java for Athene (\num{12}\%). NP and PP removals showed smaller, almost negligible effects, primarily in C and at late-sentence positions.

\smallskip
\subsubsection{Granularity}
Granularity is the weakest individual predictor, acting as a modulator rather than a primary risk driver.

Only clause-level permutations showed mild effects in Python (significant only for Qwen, RR = 1.32, p = \num{0.043}) and were associated with slightly higher vulnerability rates across languages (e.g., \num{59.72}\% in Qwen–Python, \num{12.50}\% in Qwen–Java). Minimal-level removals showed near-negligible uplift across most settings, though Athene–Java showed a significant effect (RR $\approx$ \num{1.42}, p = \num{0.011}). Chunk-level removals did not show consistent effects.
The omnibus test for Granularity was significant only for the Qwen–Python combination among all experiments, confirming that granularity in isolation does not reliably affect vulnerability incidence.\\

\steSummaryBox{\faLightbulbO\ Answer to RQ\textsubscript{1}}{The syntactic features of prompts directly affect LLMs' security performance. Risk is most sensitive to the location of the permutation (Sentence Index) and what is permuted (Constituent Type), with the extent of permutation (Granularity) playing a less impactful role. 
Java concentrates risk at the beginning of prompts, while Python shows a monotonic risk gradient toward later positions. WHNP/WHADVP constituents are the most consistently security-sensitive structures, followed by VP, SBAR, and ADJP. These effects hold across models, with differences primarily in magnitude---biggest in Python, moderate in Java, and negligible in C.}

\subsection{\textbf{RQ\textsubscript{2}} —  How do combinations of syntactic features interact to influence the security of LLM-generated code?}

We now analyze whether joint configurations of permutation characteristics produce vulnerability effects beyond those predicted by individual features. Given the large combinatorial space (up to 546 unique configurations), we focus on configurations associated with vulnerability rates above the permutation baseline and on statistically significant interactions. Complete results are available in the replication package~\cite{replication_package}.

Across these results, a consistent pattern emerges. When a single feature is associated with a higher vulnerability rate, any configuration that includes that feature also tends to show higher vulnerability than the baseline. In other words, risky constituents remain risky when combined with others, and their effect tends to accumulate rather than disappear. This behavior is consistent across all three programming languages and all three LLMs, highlighting the interaction between Position and Constituent Type as the main driver of combined risk. The main patterns emerged are summarized as follows, ordered by importance across languages.

\smallskip
\subsubsection{Early-position disruption of binding scaffolding} The most impactful combination across languages and models is [Constituent Type: WHNP/WHADVP; Sentence Index: 0, Granularity: Minimal]. In Java, this combination led to vulnerability increases of +0.27\% to +11.56\% above baseline, making it the most security-critical configuration. In Python, the same pattern yielded increases of +6.95\% to +38.75\%, reaching statistical significance across all models (RR $\approx$ 1.40–1.67). Even in C, where the vulnerability surface is sparse, combinations involving WHNP showed sensitivity with Phi-4 (+0.61\% to +0.83\%). These results confirm that disrupting the interrogative/relative scaffolding that binds entities and constraints at the prompt anchor---the sentence that concentrates core task semantics---is a cross-language risk factor.

\smallskip
\subsubsection{Late-position erosion of entity and relational constraints} The second recurrent pattern involves noun and prepositional phrase disruptions at final sentence positions: [Constituent Type: NP/PP, Late Sentence Index]. Effects appeared across all languages, ranging from +0.20\% to +2.00\% in C, +0.10\% to +9.86\% in Java, and +0.28\% to +44.23\% in Python. In Python, the positional gradient extends further: interactions involving high sentence indices (5–9) produced the most extreme effects regardless of constituent type or granularity, with vulnerability rates frequently approaching or reaching 100\% (+46.61\% to +60.90\% above baseline) and statistical significance across all models (RR $\approx$ 1.3–2.2). Together, these findings indicate that late-position constituents carry critical boundary conditions and scope markers; their removal erodes the constraints that maintain well-scoped, safe generated code.

\smallskip
\subsubsection{Removal of clausal guards and action semantics} A third pattern emerges from combinations involving SBAR and VP constituents, particularly at minimal granularity: [Constituent Type: SBAR/VP, Granularity: Minimal]. Vulnerability increases ranged from +0.07\% to +2.96\% in C and +0.14\% to +18.95\% in Java, reinforcing that conditional logic (``only if,'' ``unless'') and action semantics (``sanitize,'' ``validate'') remain critical even when the removed span is small. In Python, WHADVP, WHNP, SBAR, ADJP, and VP constituents, when combined with any sentence index or granularity level, consistently showed elevated rates (+0.81\% to +54.54\%), confirming these as the most security-impacting constituent types in prompts regardless of the accompanying configuration.

Across all three patterns, Granularity plays a secondary role. Although significant combinations exist at the Minimal, Chunk, and Clause levels, no consistent pattern indicates that any granularity level directly drives security risk. Instead, granularity modulates the effects of position and constituent type: Clause-level removals amplify interaction effects, whereas Minimal and Chunk removals produce the expected patterns without a clear independent contribution. This confirms the interpretation observed in the single-feature analysis.

The three critical patterns hold across all tested LLMs for each language: the same prompt syntax configurations raise vulnerability in the same direction across Qwen, Phi-4, and Athene. Differences across models are primarily in magnitude, not direction (i.e., no model was uniquely resistant or uniquely riskier). The pattern criticality ordering (early binding disruption $>$ late constraint erosion $>$ clausal/adverbial guard removal) holds across C, Java, and Python, indicating language- and model-independent security effects.

\steSummaryBox{\faLightbulbO\ Answer to RQ\textsubscript{2}}{Feature combinations amplify vulnerability risk: individually risky features compound when combined, and the dominant driver is Constituent Type$\times$Position. Three cross-language patterns drive this effect, in descending criticality: early-position disruption of binding scaffolding, late-position erosion of entity and relational constraints, and removal of clausal guards and action semantics. Granularity modulates rather than determines risk. These patterns are consistent across all tested LLMs, with differences only in magnitude.}
\section{Discussion and Implications}
\label{sec:Discussion}

\subsection{Discussion}


Our study confirms that prompt syntax systematically influences the security of LLM-generated code. The effect is strongest in Python, moderate in Java, and weak in C, yet directionally consistent across all three tested models. 


\smallskip
\textbf{Bidirectional fragility of prompt interpretation.}
Mirzadeh et al.~\cite{mirzadeh2024gsm} showed that \textit{adding} a single irrelevant clause can reduce GPT-4 accuracy by up to 65\%. Our results demonstrate that \textit{removing} a relevant constituent can increase vulnerability rates by comparable magnitudes (risk ratios up to 2.24). This finding suggests that LLMs do not robustly extract abstract intent but can also rely on the surface composition of the input. This implies that prompt-level security is inherently fragile and cannot be assumed to transfer across paraphrases of the same underlying request.

\textbf{Positional sensitivity and the anchor--tail duality.}
In Python, vulnerability rates increase monotonically with sentence index, approaching 100\% at the latest positions. Late-position constituents typically encode boundary conditions and operational guards; their removal strips constraints that separate secure code from merely functional code. This pattern aligns with the ``lost in the middle'' phenomenon in transformer attention~\cite{liu2024lost}: tail-positioned information is both highly attended to and highly fragile upon removal. In Java, the opposite holds: risk concentrates at Sentence Index~0, where the core task specification resides. Disrupting this \textit{anchor} destabilizes the entire generation. The two patterns reveal a duality: anchors define task semantics, and tails constrain security boundaries, and both are critical.

\textbf{Cross-model consistency.}
Despite differences in parameter count (14.7B--72B), training data, and architecture, Qwen~2.5, Athene-V2, and Phi-4 exhibit the same vulnerability patterns: the same constituent types are security-sensitive, the same positional gradients emerge, and the same Constituent Type~$\times$~Position interactions dominate. Differences are restricted to magnitude, not direction. This stability suggests that syntactic sensitivity is a fundamental property of how autoregressive models process structured instructions for code generation, rather than an artifact of any individual model.


\subsection{Implications}
In this section, we present actionable implications grounded in the quantitative evidence reported above.
The marker \faHandORight\ identifies implications for practitioners who use LLMs in their development workflows, while \faBook\ identifies implications for researchers in secure code generation and prompt engineering.

\discussionBox{\faHandORight\ \textbf{Practitioners} should always state \textit{which}, \textit{how}, and \textit{under what conditions} explicitly in the prompts.}

WHNP and WHADVP constituents, which encode phrases like ``which input'', ``how the data is validated'', or ``under what authority'', were the most consistently security-sensitive structures across all models and languages. Their removal produced the highest risk ratios (up to 1.88 in Python). These constituents bind entities to constraints. When they are missing, the model loses track of what needs to be checked, filtered, or restricted. As a consequence, practitioners should ensure that their prompts explicitly spell out which entities are involved, how they should be handled, and under what conditions operations should proceed.

\discussionBox{\faHandORight\ \textbf{Practitioners} should consider that shortening prompts can pose a security risk.}

Minimal-granularity removals (i.e., the smallest possible edits) were sufficient to increase vulnerability rates in several configurations, especially when targeting binding scaffolding (WHNP/WHADVP) at early positions. This means that even casual rewording, abbreviation, or paraphrasing of a prompt can inadvertently strip a security-relevant element. Practitioners should be cautious when simplifying or shortening prompts for convenience, and should verify that all constraint-bearing elements survive the edit.

\discussionBox{\faBook\ \textbf{Researchers} should treat prompt syntax as a primary security variable.}

Our results show that syntactic constituents are not interchangeable: their type and position produce statistically significant differences in vulnerability incidence that are consistent across models. Current secure code generation benchmarks and prompt optimization frameworks typically vary prompts at a strategic or semantic level, treating syntax as a marginal aspect. Future work should incorporate fine-grained syntactic features as explicit independent variables in experimental designs, enabling more precise identification of what makes a prompt secure or insecure.

\discussionBox{\faBook\ \textbf{Researchers} should develop syntax-aware prompt analysis tooling.}

The identification of specific constituent types (WHNP, WHADVP, SBAR, VP) and positions as security-critical control surfaces opens the door to automated, pre-generation defenses. A prompt linter that parses constituency structure and flags when security-sensitive constituents are missing, weakened, or buried in low-attention positions could serve as a lightweight complement to post-generation static analysis. Building and evaluating such tooling is a concrete next step enabled by our findings.

\section{Threats to Validity}

\textbf{Construct validity}. We identify vulnerable snippets and record the number of CWEs per snippet using CodeQL findings. While CodeQL is a state-of-the-art static analyzer used at scale, any static analysis can miss issues (false negatives) or surface non-exploitable findings (false positives), which may blur fine-grained distinctions in security outcomes. Cross-checking with dynamic analysis, symbolic execution, and consensus across independent static analyzer could be employed to strengthen validation.

\textbf{Internal validity}. Our intervention removes syntagms according to a single parser-driven permutation logic; different strategies (e.g., dependency-based or semantic role-based) could yield different effects. Although we fixed prompts and decoding settings and performed sampled reruns to confirm stability, residual nondeterminism in LLM generation may still introduce variance unrelated to the linguistic factor under test. Further experiments with alternative parsers, repeated runs, and different models are needed to assess robustness.
To ensure that our results were not influenced by randomness in the generation process, we \textbf{repeated the experiment on a statistically representative subsample} of the dataset consisting 109 baseline and 353 permutation prompts. We used the same generation and analysis pipeline, producing \textbf{3 code samples per prompt} and systematically calculating cross-runs means. The outcomes confirmed the same trends: Python remained vulnerability-prone, Java showed moderate risk, and C produced the smallest effects. Considering impactful features, the same constituents—constraint binders, threshold setters, conditional guards, action directives, identity and scope markers, and relational anchors—were those whose removal most degraded security. Moreover, the same early-sentence disruption of concepts relatives and bindings, the broad sensitivity to high-end sentence removals were also observed, and granularity modulating rather than determining risking.
These results shows that our findings are robust and not tied to a single generation. All the artifacts related to this confirmatory run are included in the replication package~\cite{replication_package}.

\textbf{Conclusion validity.} We mitigated Type~I/II error risks through (i)~a quantile-based sufficiency filter to exclude underpowered cells; (ii)~a two-stage plan---omnibus $\chi^2$ tests with Cram\'{e}r's~V, followed by Barnard's exact tests with Benjamini--Hochberg FDR control---reporting effect sizes alongside p-values; (iii)~cross-LLM and cross-language replication; and (iv)~a confirmatory subsample with multiple generations per prompt. The main residual threat concerns language-wise differences in vulnerability incidence. Python's higher rates likely reflect its broader analyzable surface (dynamic features, richer APIs, wider CodeQL coverage) and model-generated snippets that favor riskier defaults~\cite{Alfadel2023_EMSE_PythonPackages}, whereas Java and C snippets---constrained by stricter compilation and, in C's case, simpler generated patterns---exhibited lower incidence~\cite{cardoza2020_veracode_by_language, lemos2022_memory_safe_shift, qian2025softwarevulnerabilityanalysisprogramming, sakharkar2023_coding_vuln_review}. We interpret these differences as reflecting snippet complexity and analyzer coverage rather than intrinsic language security gaps. Future work will test these patterns on larger, more diverse datasets and across additional programming languages.


\textbf{External validity}. Our scenarios, prompts, and analyses target function-level completions in English across a limited set of programming languages and open models. Results may not generalize to other prompt styles, natural or programming languages, larger/proprietary systems, or project-scale contexts with tests and runtime effects. Covering additional models, languages, prompt regimes, and project-scale settings are needed to assess how broadly these findings hold.
\section{Conclusions}

Our study shows that prompt syntax directly affects the security of code produced by LLMs. The analysis highlighted how specific linguistic levers account for the vulnerability risk shifts we observe. Specifically, changing or removing constituents in the final positions, disrupting entity and constraints at the prompt anchor, and overall perturbing guards, qualifiers, or identity/scope/concept bindings made the code consistently more likely to contain security issues. 

These findings confirm syntax as a key security control surface, defined by low-level constituents which are essential for the LLM to properly understand and account users' requests and deliver secure artifacts. Prompt wording should therefore be treated as carefully as model guardrails and post-generation code checks. In future work we aim to employ different permutation techniques and account different linguistic features to further shed light on the impact of such characteristics on the security of artificially generated code.





\section*{Data Availability}
\label{sec:data:availability}
We provide a \textbf{replication package}~\cite{replication_package}, which contains the data and scripts to rerun the experiments.

\balance
\bibliographystyle{IEEEtran}
\bibliography{references}

@article{chen2021evaluating,
  title={Evaluating large language models trained on code},
  author={Chen, Mark and Tworek, Jerry and Jun, Heewoo and Yuan, Qiming and Pinto, Henrique Ponde De Oliveira and Kaplan, Jared and Edwards, Harri and Burda, Yuri and Joseph, Nicholas and Brockman, Greg and others},
  journal={arXiv preprint arXiv:2107.03374},
  year={2021}
}

@article{pearce2025asleep,
  title={Asleep at the keyboard? assessing the security of github copilot’s code contributions},
  author={Pearce, Hammond and Ahmad, Baleegh and Tan, Benjamin and Dolan-Gavitt, Brendan and Karri, Ramesh},
  journal={Communications of the ACM},
  volume={68},
  number={2},
  pages={96--105},
  year={2025},
  publisher={ACM New York, NY, USA}
}

@inproceedings{perry2023users,
  title={Do users write more insecure code with ai assistants?},
  author={Perry, Neil and Srivastava, Megha and Kumar, Deepak and Boneh, Dan},
  booktitle={Proceedings of the 2023 ACM SIGSAC conference on computer and communications security},
  pages={2785--2799},
  year={2023}
}

@inproceedings{bruni2025benchmarking,
  title={Benchmarking prompt engineering techniques for secure code generation with gpt models},
  author={Bruni, Marc and Gabrielli, Fabio and Ghafari, Mohammad and Kropp, Martin},
  booktitle={2025 IEEE/ACM Second International Conference on AI Foundation Models and Software Engineering (Forge)},
  pages={93--103},
  year={2025},
  organization={IEEE}
}

@inproceedings{tony2023llmseceval,
  title={Llmseceval: A dataset of natural language prompts for security evaluations},
  author={Tony, Catherine and Mutas, Markus and Ferreyra, Nicol{\'a}s E D{\'\i}az and Scandariato, Riccardo},
  booktitle={2023 IEEE/ACM 20th International Conference on Mining Software Repositories (MSR)},
  pages={588--592},
  year={2023},
  organization={IEEE}
}

@article{tony2025prompting,
  title={Prompting techniques for secure code generation: A systematic investigation},
  author={Tony, Catherine and D{\'\i}az Ferreyra, Nicol{\'a}s E and Mutas, Markus and Dhif, Salem and Scandariato, Riccardo},
  journal={ACM Transactions on Software Engineering and Methodology},
  volume={34},
  number={8},
  pages={1--53},
  year={2025},
  publisher={ACM New York, NY}
}

@INPROCEEDINGS{tony2025:icsme:rag,
  author={Tony, Catherine and Iannone, Emanuele and Scandariato, Riccardo},
  booktitle={2025 IEEE International Conference on Software Maintenance and Evolution (ICSME)}, 
  title={Retrieve, Refine, or Both? Using Task-Specific Guidelines for Secure Python Code Generation}, 
  year={2025},
  volume={},
  number={},
  pages={368-379},
  doi={10.1109/ICSME64153.2025.00041}}

@inproceedings{prasad2023grips,
  title={Grips: Gradient-free, edit-based instruction search for prompting large language models},
  author={Prasad, Archiki and Hase, Peter and Zhou, Xiang and Bansal, Mohit},
  booktitle={Proceedings of the 17th Conference of the European Chapter of the Association for Computational Linguistics},
  pages={3845--3864},
  year={2023}
}

@article{tian2024selective,
  title={Selective prompt anchoring for code generation},
  author={Tian, Yuan and Zhang, Tianyi},
  journal={arXiv preprint arXiv:2408.09121},
  year={2024}
}

@article{chen2024nlperturbator,
  title={Nlperturbator: Studying the robustness of code llms to natural language variations},
  author={Chen, Junkai and Zhenhao, Li and Xing, Hu and Xin, Xia},
  journal={ACM Transactions on Software Engineering and Methodology},
  year={2024},
  publisher={ACM New York, NY}
}

@article{paleyes2025prompt,
  title={Prompt variability effects on LLM code generation},
  author={Paleyes, Andrei and Sendyka, Radzim and Robinson, Diana and Cabrera, Christian and Lawrence, Neil D},
  journal={arXiv preprint arXiv:2506.10204},
  year={2025}
}

@article{viveros2025does,
  title={Does the grammatical structure of prompts influence the responses of generative artificial intelligence? An exploratory analysis in Spanish},
  author={Viveros-Mu{\~n}oz, Rhoddy and Carrasco-S{\'a}ez, Jos{\'e} and Contreras-Saavedra, Carolina and San-Mart{\'\i}n-Quiroga, Sheny and Contreras-Saavedra, Carla E},
  journal={Applied Sciences},
  volume={15},
  number={7},
  pages={3882},
  year={2025},
  publisher={MDPI}
}

@article{mirzadeh2024gsm,
  title={Gsm-symbolic: Understanding the limitations of mathematical reasoning in large language models},
  author={Mirzadeh, Iman and Alizadeh, Keivan and Shahrokhi, Hooman and Tuzel, Oncel and Bengio, Samy and Farajtabar, Mehrdad},
  journal={arXiv preprint arXiv:2410.05229},
  year={2024}
}

@article{guo2023connecting,
  title={Connecting large language models with evolutionary algorithms yields powerful prompt optimizers},
  author={Guo, Qingyan and Wang, Rui and Guo, Junliang and Li, Bei and Song, Kaitao and Tan, Xu and Liu, Guoqing and Bian, Jiang and Yang, Yujiu},
  journal={arXiv preprint arXiv:2309.08532},
  year={2023}
}

@inproceedings{siddiq2024sallm,
  title={Sallm: Security assessment of generated code},
  author={Siddiq, Mohammed Latif and da Silva Santos, Joanna Cecilia and Devareddy, Sajith and Muller, Anna},
  booktitle={Proceedings of the 39th IEEE/ACM International Conference on Automated Software Engineering Workshops},
  pages={54--65},
  year={2024}
}

@inproceedings{xiao2024devgpt,
  title={Devgpt: Studying developer-chatgpt conversations},
  author={Xiao, Tao and Treude, Christoph and Hata, Hideaki and Matsumoto, Kenichi},
  booktitle={Proceedings of the 21st international conference on mining software repositories},
  pages={227--230},
  year={2024}
}

@inproceedings{della2025prompt,
  title={Do prompt patterns affect code quality? a first empirical assessment of chatgpt-generated code},
  author={Della Porta, Antonio and Lambiase, Stefano and Palomba, Fabio},
  booktitle={Proceedings of the 29th International Conference on Evaluation and Assessment in Software Engineering},
  pages={181--192},
  year={2025}
}

@online{Penn_Treebanks, 
url={https://surdeanu.cs.arizona.edu/mihai/teaching/ista555-fall13/readings/PennTreebankConstituents.html}, journal={Penn Treebank Constituent Tags}}

@article{marcus-etal-1993-building,
    title = "Building a Large Annotated Corpus of {E}nglish: The {P}enn {T}reebank",
    author = "Marcus, Mitchell P.  and
      Santorini, Beatrice  and
      Marcinkiewicz, Mary Ann",
    editor = "Hirschberg, Julia",
    journal = "Computational Linguistics",
    volume = "19",
    number = "2",
    year = "1993",
    address = "Cambridge, MA",
    publisher = "MIT Press",
    pages = "313--330"
}

@article{sakharkar2023_coding_vuln_review,
  author  = {Shreyas Sakharkar},
  title   = {Systematic Review: Analysis of Coding Vulnerabilities across Languages},
  journal = {Journal of Information Security},
  year    = {2023},
  volume  = {14},
  pages   = {330--342},
  doi     = {10.4236/jis.2023.144019},
}

@misc{qian2025softwarevulnerabilityanalysisprogramming,
      title={Software Vulnerability Analysis Across Programming Language and Program Representation Landscapes: A Survey}, 
      author={Zhuoyun Qian and Fangtian Zhong and Qin Hu and Yili Jiang and Jiaqi Huang and Mengfei Ren and Jiguo Yu},
      year={2025},
      eprint={2503.20244},
      archivePrefix={arXiv},
      primaryClass={cs.CR},
}

@misc{lemos2022_memory_safe_shift,
  author       = {Robert Lemos},
  title        = {Shift to Memory-Safe Languages Gains Momentum},
  howpublished = {Dark Reading},
  year         = {2022},
  month        = {December},
  day          = {7},
  url          = {https://www.darkreading.com/application-security/shift-memory-safe-languages-gains-momentum},
  note         = {Accessed 2025-10-01}
}

@misc{cardoza2020_veracode_by_language,
  author       = {Christina Cardoza},
  title        = {Veracode Uncovers the Top Security Issues Facing Specific Programming Languages},
  howpublished = {SD Times},
  year         = {2020},
  month        = {December},
  day          = {15},
  note         = {Accessed 2025-10-01}
}

@article{Alfadel2023_EMSE_PythonPackages,
  author  = {Alfadel, Mahmoud and Costa, Diego Elias and Shihab, Emad},
  title   = {Empirical analysis of security vulnerabilities in Python packages},
  journal = {Empirical Software Engineering},
  year    = {2023},
  volume  = {28},
  number  = {3},
  pages   = {59},
  doi     = {10.1007/s10664-022-10278-4},
  url     = {https://dl.acm.org/doi/10.1007/s10664-022-10278-4}
}

@misc{replication_package,
    author = {Cicalese, M and Della Porta, A. and Lambiase, S. and Iannone, E. and Hinrichs, T and Scandariato, R. and Palomba, F.},
    howpublished = {\url{https://doi.org/10.6084/m9.figshare.32707911}}
}

@inproceedings{hajipour2024codelmsec,
  title={CodeLMSec benchmark: Systematically evaluating and finding security vulnerabilities in black-box code language models},
  author={Hajipour, Hossein and Hassler, Keno and Holz, Thorsten and Sch{\"o}nherr, Lea and Fritz, Mario},
  booktitle={2024 IEEE Conference on Secure and Trustworthy Machine Learning (SaTML)},
  pages={684--709},
  year={2024},
  organization={IEEE}
}

@article{jiang2024survey,
  title={A survey on large language models for code generation},
  author={Jiang, Juyong and Wang, Fan and Shen, Jiasi and Kim, Sungju and Kim, Sunghun},
  journal={ACM Transactions on Software Engineering and Methodology},
  year={2024},
  publisher={ACM New York, NY}
}

@inproceedings{hamer2024just,
  title={Just another copy and paste? comparing the security vulnerabilities of chatgpt generated code and stackoverflow answers},
  author={Hamer, Sivana and d’Amorim, Marcelo and Williams, Laurie},
  booktitle={2024 IEEE Security and Privacy Workshops (SPW)},
  pages={87--94},
  year={2024},
  organization={IEEE}
}

@article{li2024exploratory,
  title={An exploratory study on fine-tuning large language models for secure code generation},
  author={Li, Junjie and Rabbi, Fazle and Cheng, Cheng and Sangalay, Aseem and Tian, Yuan and Yang, Jinqiu},
  journal={arXiv preprint arXiv:2408.09078},
  year={2024}
}

@inproceedings{zhang2020fastcrfconst,
  title     = {Fast and Accurate Neural CRF Constituency Parsing},
  author    = {Zhang, Yu and Zhou, Houquan and Li, Zhenghua},
  booktitle = {Proceedings of the 29th International Joint Conference on Artificial Intelligence (IJCAI)},
  year      = {2020},
  pages     = {4044--4051},
  url       = {https://www.ijcai.org/proceedings/2020/560}
}

@book{wohlin2012experimentation,
  title={Experimentation in software engineering},
  author={Wohlin, Claes and Runeson, Per and H{\"o}st, Martin and Ohlsson, Magnus C and Regnell, Bj{\"o}rn and Wessl{\'e}n, Anders and others},
  volume={236},
  year={2012},
  publisher={Springer}
}

@inproceedings{prasad2023gradient,
  title={Grips: Gradient-free, edit-based instruction search for prompting large language models},
  author={Prasad, Archiki and Hase, Peter and Zhou, Xiang and Bansal, Mohit},
  booktitle={Proceedings of the 17th Conference of the European Chapter of the Association for Computational Linguistics},
  pages={3845--3864},
  year={2023}
}

@article{liao2025hierarchical,
  title={Hierarchical fine-grained state-aware graph attention network for dialogue state tracking},
  author={Liao, Hongmiao and Chen, Yuzhong and Chen, Deming and Xu, Junjie and Zhong, Jiayuan and Dong, Chen},
  journal={The Journal of Supercomputing},
  volume={81},
  number={5},
  pages={671},
  year={2025},
  publisher={Springer}
}

@article{wu2024knowledge,
  title={Knowledge Graph-Based Hierarchical Text Semantic Representation},
  author={Wu, Yongliang and Pan, Xiao and Li, Jinghui and Dou, Shimao and Dong, Jiahao and Wei, Dan},
  journal={International journal of intelligent systems},
  volume={2024},
  number={1},
  pages={5583270},
  year={2024},
  publisher={Wiley Online Library}
}

@inproceedings{wu2022text,
  title={Text Semantic Representation Based on Knowledge Graph Correction},
  author={Wu, Yongliang and Yin, Hu and Liu, Dongbo and Zhou, Qianqian},
  booktitle={2022 International Conference on Computer Engineering and Artificial Intelligence (ICCEAI)},
  pages={404--408},
  year={2022},
  organization={IEEE}
}

@inproceedings{wu2022answer,
  title={An Answer Recommendation Algorithm Based on Semantic Fusion Heterogeneous Information Network},
  author={Wu, Yongliang and Liu, Dongbo and Zhou, Qianqian and Yin, Hu},
  booktitle={2022 International Conference on Computer Engineering and Artificial Intelligence (ICCEAI)},
  pages={63--67},
  year={2022},
  organization={IEEE}
}

@inproceedings{schreiber2025security,
  title={Security Vulnerabilities in AI-Generated Code: A Large-Scale Analysis of Public GitHub Repositories},
  author={Schreiber, Maximilian and Tippe, Pascal},
  booktitle={International Conference on Information and Communications Security},
  pages={153--172},
  year={2025},
  organization={Springer}
}

@inproceedings{shukla2025security,
  title={Security degradation in iterative AI code generation: A systematic analysis of the paradox},
  author={Shukla, Shivani and Joshi, Himanshu and Syed, Romilla},
  booktitle={2025 IEEE International Symposium on Technology and Society (ISTAS)},
  pages={1--8},
  year={2025},
  organization={IEEE}
}

@inproceedings{cotroneo2025human,
  title={Human-written vs. ai-generated code: A large-scale study of defects, vulnerabilities, and complexity},
  author={Cotroneo, Domenico and Improta, Cristina and Liguori, Pietro},
  booktitle={2025 IEEE 36th International Symposium on Software Reliability Engineering (ISSRE)},
  pages={252--263},
  year={2025},
  organization={IEEE}
}

@article{bommasani2021opportunities,
  title={On the opportunities and risks of foundation models},
  author={Bommasani, Rishi and Hudson, Drew A and Adeli, Ehsan and Altman, Russ and Arora, Simran and von Arx, Sydney and Bernstein, Michael S and Bohg, Jeannette and Bosselut, Antoine and Brunskill, Emma and others},
  journal={arXiv preprint arXiv:2108.07258},
  year={2021}
}

@article{das2025security,
  title={Security and privacy challenges of large language models: A survey},
  author={Das, Badhan Chandra and Amini, M Hadi and Wu, Yanzhao},
  journal={ACM Computing Surveys},
  volume={57},
  number={6},
  pages={1--39},
  year={2025},
  publisher={ACM New York, NY}
}

@article{liu2024lost,
  title={Lost in the middle: How language models use long contexts},
  author={Liu, Nelson F and Lin, Kevin and Hewitt, John and Paranjape, Ashwin and Bevilacqua, Michele and Petroni, Fabio and Liang, Percy},
  journal={Transactions of the association for computational linguistics},
  volume={12},
  pages={157--173},
  year={2024}
}

@inproceedings{della2025unlocking,
  title={Unlocking code simplicity: The role of prompt patterns in managing llm code complexity},
  author={Della Porta, Antonio and Recupito, Gilberto and Lambiase, Stefano and Di Nucci, Dario and Palomba, Fabio},
  booktitle={2025 IEEE International Conference on Software Analysis, Evolution and Reengineering-Companion (SANER-C)},
  pages={140--143},
  year={2025},
  organization={IEEE}
}

@inproceedings{della2026toward,
  title={Toward Measuring Prompt Quality: A Preliminary Investigation on Prompt Smells},
  author={Della Porta, Antonio and Voria, Gianmario and Abbate, Andrea and Sulipano, Raffaele and Lambiase, Stefano and Catolino, Gemma and Palomba, Fabio},
  booktitle={2026 IEEE International Conference on Software Analysis, Evolution and Reengineering-Companion (SANER-C)},
  pages={293--300},
  year={2026},
  organization={IEEE}
}

@article{cannavale2025fairness,
  title={Fairness set and forgotten: Mining fairness toolkit usage in open-source machine learning projects},
  author={Cannavale, Alfonso and Voria, Gianmario and Scognamiglio, Antonio and Giordano, Giammaria and Catolino, Gemma and Palomba, Fabio},
  journal={Information and Software Technology},
  pages={107957},
  year={2025},
  publisher={Elsevier}
}

@inproceedings{parziale2025contextual,
  title={Contextual fairness-aware practices in ML: A cost-effective empirical evaluation},
  author={Parziale, Alessandra and Voria, Gianmario and Giordano, Giammaria and Catolino, Gemma and Robles, Gregorio and Palomba, Fabio},
  booktitle={2025 IEEE International Conference on Software Analysis, Evolution and Reengineering-Companion (SANER-C)},
  pages={1--8},
  year={2025},
  organization={IEEE}
}

@article{parziale2025fairness,
  title={Fairness on a budget, across the board: A cost-effective evaluation of fairness-aware practices across contexts, tasks, and sensitive attributes},
  author={Parziale, Alessandra and Voria, Gianmario and Giordano, Giammaria and Catolino, Gemma and Robles, Gregorio and Palomba, Fabio},
  journal={Information and Software Technology},
  pages={107858},
  year={2025},
  publisher={Elsevier}
}

@article{voria2025fair,
  title={Fair and square? Evaluating fairness of LLM-generated synthetic datasets},
  author={Voria, Gianmario and Scala, Benedetto and Todisco, Leopoldo and Venditto, Carlo and Giordano, Giammaria and Catolino, Gemma and Palomba, Fabio},
  journal={Information and Software Technology},
  pages={107980},
  year={2025},
  publisher={Elsevier}
}

@article{recupito2025code,
  title={When code smells meet ML: on the lifecycle of ML-specific code smells in ML-enabled systems},
  author={Recupito, Gilberto and Giordano, Giammaria and Ferrucci, Filomena and Di Nucci, Dario and Palomba, Fabio},
  journal={Empirical Software Engineering},
  volume={30},
  number={5},
  pages={139},
  year={2025},
  publisher={Springer}
}

@article{de2025into,
  title={Into the ml-universe: An improved classification and characterization of machine-learning projects},
  author={De Martino, Vincenzo and Recupito, Gilberto and Giordano, Giammaria and Ferrucci, Filomena and Di Nucci, Dario and Palomba, Fabio},
  journal={Journal of Systems and Software},
  volume={230},
  pages={112471},
  year={2025},
  publisher={Elsevier}
}

@inproceedings{giordano2025evidence,
  title={An evidence-based study on the relationship of software engineering practices on code smells in python ml projects},
  author={Giordano, Giammaria and Della Porta, Antonio and Ferrucci, Filomena and Palomba, Fabio},
  booktitle={Euromicro Conference on Software Engineering and Advanced Applications},
  pages={105--120},
  year={2025},
  organization={Springer}
}

@article{della2024using,
  title={Using large language models to support software engineering documentation in waterfall life cycles: Are we there yet?},
  author={Della Porta, Antonio and De Martino, Vincenzo and Recupito, Gilberto and Iemmino, Carmine and Catolino, Gemma and Di Nucci, Dario and Palomba, Fabio and others},
  journal={Ital-IA},
  pages={42--47},
  year={2024}
}

@inproceedings{giordano2023understanding,
  title={Understanding Developer Practices and Code Smells Diffusion in AI-Enabled Software: A Preliminary Study.},
  author={Giordano, Giammaria and Annunziata, Giusy and De Lucia, Andrea and Palomba, Fabio and others},
  booktitle={IWSM-Mensura},
  year={2023}
}

\end{document}